# Ownership Cost Calculations for Distributed Energy Resources Using Uncertainty and Risk Analyses

S. Ali Pourmousavi, Mahdi Behrangrad, Ali Jahanbani Ardakani, and M. Hashem Nehrir

*Abstract*—Ownership cost calculation plays an important role in optimal operation of distributed energy resources (DERs) and microgrids (MGs) in the future power system, known as smart grid. In this paper, a general framework for ownership cost calculation is proposed using uncertainty and risk analyses. Four ownership cost calculation approaches are introduced and compared based on their associated risk values. Finally, the best method is chosen based on a series of simulation results, performed for a typical diesel generator (DiG). Although simulation results are given for a DiG (as commonly used in MGs), the proposed approaches can be applied to other MG components, such as batteries, with slight modifications, as presented in this paper. The analyses and proposed approaches can be useful in MG optimal design, optimal power flow, and market-based operation of the smart grid for accurate operational cost calculations.

*Index Terms*—Battery, diesel generator (DiG), distributed energy resources (DERs), expected value, ownership cost, risk analysis, uncertainty.

## I. INTRODUCTION

IN the smart grid era, cost-based operation of distributed energy resources (DERs) and microgrids (MGs), including their ownership costs, is inevitable. Strictly speaking, accurate operational cost calculation plays an important role for profitable operation of DERs for their owners, as one of the major goals projected for the smart grid [1]. In general, different uncertainty sources can be envisioned for different elements of the operational cost (such as depreciation and operating costs), which are the actual causes of inaccuracy in the cost calculations. Therefore, an accurate operational cost calculation model considering uncertainties with lower risk is essential for power/energy management and optimal sizing of MGs, as well as for optimal power flow, power scheduling, and market-based operation of power systems. This paper proposes a general framework for DERs ownership cost calculation, which is an important part of the operational cost considering existing uncertainties and associated risks.

Power and energy management of islanded and grid-tied MGs is important for their optimal operation and to minimize their operational costs, e.g. [2]-[19]. However in these references, the ownership cost of the equipment is not investigated explicitly nor in some cases considered in the optimal MG operation. For example, only fuel cost is considered for a diesel generator (DiG), [2], [3], [5]-[9], [13], [15]; or in some other literature, e.g., [11], [12], [16], [17], the cost of DiG operation is considered as a price offered by the owner (similar to bids given by the participants in the electricity market). Also in [4] and [10], the battery cost is not included in the proposed management algorithm. In [14] and [19], the battery cost is explored based on the estimated battery lifetime, however the proposed approaches do not consider the uncertainties in the operation of the battery that could alter the true battery cost. As a result, substantial fluctuations were observed in the estimated cost, particularly in the earliest days of operation, when a high level of uncertainty occurs. In [18], the battery ownership cost is incorporated in the MG operation objective function, however it will be shown in this study that it will not always give accurate results.

There is a great amount of literature in the area of optimal sizing of MG components, where cost of all equipment (such as DiG and battery) is considered. A comprehensive literature review on this topic is given in [20]. Nevertheless, the actual operational uncertainties and associated risks in the ownership cost calculations are not investigated. Similar deficiencies exist in several technical reports on the cost-benefit analysis of battery technologies, e.g. [21], [22]. All in all, there is no independent study on the ownership cost calculation, to the best of our knowledge.

This paper lays out a general framework for DERs ownership cost calculation considering uncertainty and risk analyses. Initially, four different approaches are presented to calculate DERs ownership cost, where two of the approaches are proposed by the authors. Various uncertainty sources in the calculations are then introduced, where an appropriate discrete probability density function (DPDF) is specified for each in the simulation studies. In order to define the appropriate DPDFs for the uncertain parameters, an islanded

S.A. Pourmousavi (a.pour@uq.edu.au) is with the Global Change Institute, University of Queensland, St Lucia 4072 Australia; Mahdi Behrang (m.behrang.r@gmail.com) is with the Sumitomo Electric Industries, Ltd., Osaka, Japan; Ali Jahanbani Ardakani (alij@iastate.edu) is with the Department of Electrical \& Computer Engineering, Iowa State University, Ames, Iowa, USA; and M.H. Nehrir (email: hnehrir@ece.montana.edu) are with the Electrical and Computer Engineering Department, Montana State University, Bozeman, 59717 USA,.

MG has been designed using HOMER® [23] and the data obtained (given in Section IV) has been utilized. Then, a general basis for sensitivity and risk analyses based on the DPDFs is developed. Finally, the best model is chosen based on the uncertainty analysis along with risk calculation through simulation studies. To show the applicability and adaptability of the proposed framework to other DERs, the proposed model is also adapted for battery ownership cost calculations.

The rest of the paper is organized as follows. The problem statement is presented in Section II. The proposed general framework and four different approaches to solve the ownership cost calculation problem along with a framework for uncertainty and risk analyses are given in Section III. Section IV presents several simulation studies to verify the validity of the proposed approaches and assess them. The proposed framework and required modification for battery ownership cost calculations are discussed in Section V. The conclusion is given in Section VI.

## II. PROBLEM STATEMENT

The operational cost calculation (known as machinery cost in industrial engineering and engineering economic analysis) is typically divided into two categories (as shown in Fig. 1): 1) "*ownership costs*", which occur regardless of machine use (usually expressed per hour of operation), and 2) "*operating costs*", which vary directly with the amount of machine use [24], [25]. The true value of these costs cannot be known until the machine is sold or worn out. However, these costs can be estimated by making a few assumptions about machines' life, annual use, and fuel and labor costs. These assumptions should be treated as uncertainty sources in the calculations through uncertainty and risk analyses.

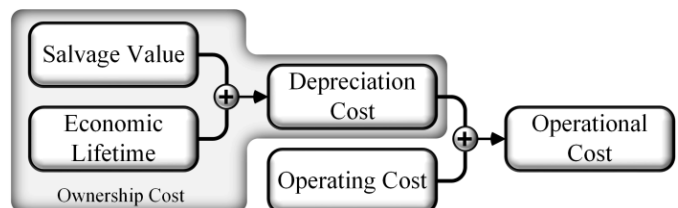

Fig. 1. Overview of the operational cost calculation model adapted for this study.

Ownership costs (also called fixed costs) include depreciation, interest (opportunity cost), taxes, insurance, and housing and maintenance facilities [24], [25]. In MGs, it is not common to insure the lifetime of a device. Therefore, insurance (insurance of equipment replacement in case of fire or other natural disaster) is neglected [26]. The housing cost (the cost of shelter or building to keep the equipment safe) is included in the capital investment of the equipment. However, property taxes which are commonly considered in farm machinery cost estimations [26], are ignored in our study, as they are usually less than 1% of the purchase cost altogether. Instead, we will focus on the most important parameters, such as machines' lifetime and their annual use. Therefore, the ownership cost calculation will be limited to the depreciation cost calculation and its components (i.e., salvage value and economic lifetime).

The *operating cost* is the second and last part of the operational cost calculations. This cost can be so different from one DER to another; even from one manufacturer to another; and from one size to another. Therefore, it is complicated, if not impossible, to develop a general framework for operating cost calculation for different DERs. Furthermore, the operating cost is significantly less than the ownership cost for many DERs, such as battery, solar Photovoltaic (PV) panels, wind turbines, etc. As an example, the operating cost of solar PV panels is approximated as 0.005 $/kW in [4], which is negligible compared to the operating cost of DiG. For DiGs, the operating cost is an important factor since it includes fuel cost. This paper only explores ownership cost calculations considering uncertainties.

## III. THE PROPOSED GENERAL FRAMEWORK FOR OWNERSHIP COST CALCULATION

Depreciation is a cost subsequent from wear, obsolescence, and age of a machine where age and accumulated hours of use are usually the most important factors in determining the remaining value of a machine [24]-[26]. Prior to annual depreciation cost calculations, *economic life* for the machine and the *salvage value* at the end of its economic lifetime are necessary to be specified.

*Economic Life*

There are two different economic lifetimes considered in this study: 1) The *economic life of an equipment*, defined as the number of years (for PVs), hours (for DiGs), or ampere-hours (A.h.) for battery, for which costs are to be approximated, and 2) The *project lifetime*, which can be expressed as the economic lifetime of a project based on the longest equipment lifetime in the system, or other considerations such as the user's expected lifetime of the whole project. As a result, some of the equipment with relatively shorter lifetime, compared to the project lifetime, will experience replacement(s) during the project lifetime. For instance, if a DiG lasts six years, and the project lifetime is expected to be 20 years, the DiG will have to be replaced three times throughout the project lifetime (i.e., at year six, 12 and 18 of the project).

Although the useful lifetime of different devices are expressed in different terms, similar principles can be employed for their ownership cost calculation with minor modifications.

*Salvage Value*

The *salvage value* is an estimate of the sale value of the equipment at the end of its economic lifetime [24]-[26]. As it is shown in Fig. 1, a prior knowledge about this value is essential at the time of ownership cost calculation. Without loss of generality, this value is considered to be zero in this study because equipment such as DiGs and batteries will be used until they are totally worn out. As a result, they are assumed to have no value at the end of their lifetime.

Based on the definition and assumptions given above, four different approaches are introduced below for ownership cost calculation. Then, uncertainty sources are identified and a method for performing risk analysis is developed. Here, all equations are developed for DiG without loss of generality. Simulation results and comparisons of the four approaches are given in Section IV. Similar equations with slight modifications (as given for battery in Section V) can be utilized for other DERs.

*A. Approach I (Base-case): No Replacement Cost*

In this approach, the depreciation cost is calculated based on the equipment's economic lifetime and its salvage value. It is also assumed that the lifetime of the project is the same as the economic lifetime of the equipment. This is a standard approach in the ownership cost calculation [24] and will be considered as the base-case scenario for comparison with other approaches.

The depreciation cost ($C_{dep}$) can be expressed as:

$$C_{dep} = C_{pur} - C_{sal} \quad (1)$$

where $C_{pur}$ and $C_{sal}$ are the capital cost and salvage value of the equipment ($), respectively. Since no replacement of the equipment is considered throughout the project lifetime:

$$C_{pur} = C_{pur,0} \quad (2)$$

where $C_{pur,0}$ ($) is the current purchase (capital) price of the equipment, including installation fees and transportation. The combined costs of depreciation and interest can be calculated by using the *capital recovery factor*. This cost is called equivalent uniform annual cost (EUAC) [25]. Capital recovery is the dollar amount that would have to be set aside each year to repay the value lost due to depreciation, and pay the interest costs [25], as given below:

$$C_{EUAC} = C_{dep} \times CRF(i,n) \quad (3)$$

where $CRF(i,n)$ is:

$$CRF(i,n) = \frac{i \times (1+i)^n}{(1+i)^n - 1} \quad (4)$$

$n$ is the project lifetime, and $i$ is the effective annual interest rate calculated in terms of the effective annual interest rate ($i_{int}$) and inflation ($i_{inf}$) as follows:

$$i = \frac{i_{int} - i_{inf}}{1 + i_{inf}} \quad (5)$$

$i_{int}$ can be considered as opportunity cost of investment if someone uses its own capital and $i_{inf}$ is the annual inflation rate. This method of depreciation calculation is called *Straight-line depreciation* [25]. Effective interest rate is considered to be constant throughout the lifetime of the project since its variation does not significantly affect the calculations.

The ownership cost per hour for the DiG can be estimated by estimating the annual operation hours of the DiG, neglecting insurance, taxes and housing costs of the equipment, as follows [26]:

$$C_I^h = \frac{C_{EUAC}}{h_y} \quad (6)$$

where $h_y$ is the estimated annual hours of operation for the DiG studied.

*B. Approach II: No Replacement Cost and Effective Annual Interest Rate*

In this approach, which has been widely used in literature, e.g., [18], the effective interest rate and equipment replacement are ignored. The hourly ownership cost of DiG (per A.h. in case of batteries) can be determined by (7):

$$C_{II}^h = \frac{C_{dep}}{h_y} \quad (7)$$

where $C_{dep}$ is calculated by (1), and $h_n$ is the estimated economic lifetime of the equipment (in terms of hours for DiGs and A.h. for batteries). It will be revealed in the simulation studies that Approach II yields erroneous results which could lead to a series of incorrect decisions, e.g., in the cost-based power/energy management algorithms or in the electricity market.

*C. Approach III: Cost Calculations Considering Replacement Cost*

In this proposed approach, the purchase cost of the equipment, defined by (2), will be modified to include replacement(s) cost as follows:

$$C_{pur} = C_{pur,0} + C_{rep} \times NO_{rep} \quad (8)$$

where $C_{rep}$ is the replacement cost of the equipment; and $NO_{rep}$ is the number of replacements of the equipment throughout the project lifetime. Replacement cost is considered to be 70% of the present worth (i.e., the current purchase price), since some installation fees such as wiring or housing will not be paid again. Here, we propose two approaches (III.A and III.B explained below) to calculate the ownership cost using the depreciation cost explained above and considering replacement cost.

*Approach III.A-EUAC with replacement:* In this approach, a similar procedure to Approach I will be executed using (3)-(5), to calculate the EUAC. Finally, the ownership cost per hour for the DiG can be calculated using the estimated annual operation hours of the DiG ($h_y$) as follows:

$$C_{III.A}^h = \frac{C_{EUAC}}{h_y} \quad (9)$$

It can be seen from (9) that the annual operation hours of the equipment is utilized to normalize the ownership cost. As it can be noticed from the explanation given above, so far the difference between Approach I and Approach III.A is only the depreciation cost calculation considering replacement costs.

*Approach III.B-Depreciation with replacement:* Similar to Approach III.A, depreciation cost of the equipment will be calculated using the replacement cost of the equipment throughout the lifetime of the project. However, the EUAC is not used and depreciation cost will be directly utilized for ownership cost calculation. Contrary to Approach III.A, in Approach III.B, the total depreciation cost is divided by the estimated economic lifetime of the equipment (expressed in terms of hours for DiG, and in A.h. for batteries) to calculate the cost per hour for DiG (or per A.h. for battery) as follows:

$$C_{III.B}^{h} = \frac{C_{dep}}{h_n \cdot (NO_{rep}+1)} \times (1+i)^{j} \qquad (10)$$

where $C_{dep}$ is calculated by (8); and $j$ is the current year of operation.

From the equations developed for the different Approaches, different sources of uncertainty can be identified for the equipment ownership cost calculation. It basically implies that the ownership cost calculations may not be accurate unless the variations of these parameters are considered, further meaning that uncertainty and risk analyses are required for accurate calculations. In addition, risk analysis provides a basis to compare the effectiveness and accuracy of the given Approaches. The uncertainty sources and corresponding uncertainty and risk analyses are presented in the next sub-section.

*D. Uncertainty and Risk Analysis:*

Several uncertainty sources can be identified as follows:
1) Equipment's number of replacement throughout the project lifetime (which depends on its economic lifetime and annual operation pattern),
2) Equipment's cost of replacement,
3) Inflation and interest rates.

Equipment's cost of replacement is highly unpredictable for new technologies since any breakthrough in a technology or unexpected market changes can influence the cost of the technology. For mature technologies such as DiGs, it is less likely to see any significant changes in the cost. Therefore, equipment's cost of replacement is not included as an uncertainty source for DiG. Also, inflation and interest rates are very stable in developed countries and do not present substantial variations. In this study, as the first attempt to consider uncertainty in the ownership cost calculation of MG components, attention is put on the first source of uncertainty, which can significantly affect the ownership cost, i.e. uncertainty source 1. The uncertainty sources 2 and 3 are neglected in this paper. However, these uncertainties can be included in the calculations if desired.

The number of replacement of an equipment ($NO_{rep}$) is primarily related to the usage pattern of the equipment and can be different from one year to another. It is also a function of the economic lifetime of the equipment which is not a certain parameter either, as discussed earlier. Although reasonable values can be assigned to the annual operation hours and economic lifetime of the equipment to obtain its number of replacement(s), they will be treated as uncertainty sources with appropriate DPDF for uncertainty and risk analyses in this study.

It is common in engineering economic analysis to use two to five outcomes with discrete probabilities to assess the uncertainty associated with the future events (such as equipment's annual operation hours and economic lifetime) [24], [25]. In most cases, the two to five outcomes represent the best trade-off between representing the range of possibilities and the amount of calculation required [25]. This is usually based on expert judgment or through prior experimental or simulation studies. Therefore, appropriate DPDF should be assigned to each uncertain parameter, given in Section IV, prior to the risk analysis.

With the values estimated by the chosen DPDFs for the two sources of uncertainty (i.e., equipment's annual operation hours and useful lifetime), it is conceivable to calculate the hourly cost for each possible outcome. It essentially yields the sensitivity of the ownership cost with respect to the uncertain parameters. Although the impact of uncertain parameters on the ownership cost can be observed in the sensitivity analysis, this analysis is not efficient for larger number of uncertainty sources and outcomes. Risk calculation can be very helpful in this context; it provides a legitimate basis to compare different approaches to find the most robust and reliable one.

As defined in [25], risk can be thought of as the chance of getting an outcome other than the expected value with an emphasis on something unpredictable. The common risk measure is the standard deviation, which measures the dispersion of outcomes about the expected value. Prior to risk calculation, it is required to calculate the expected values of each outcome for both uncertainty sources. It is therefore essential to calculate the joint probability distributions since there are two different uncertainties. It is assumed that the two uncertain parameters chosen are statistically independent, where the joint probability of a combined event is simply the product of the probabilities for the two events [25]. It is intuitive assumption that the annual operation hours of a DiG has no meaningful impact on its useful lifetime (i.e., the total operation hours) or vice versa. Therefore, the expected value of hourly cost for each approach can be drawn from (11) [25]:

$$EV_i\left[C_i^h\right] = \sum_{m=1}^{ELT}\sum_{n=1}^{AOH}\left\langle P_{\{X=m\}}^{ELT} \cdot P_{\{X=n\}}^{AOH} \cdot C_i^h(m,n)\right\rangle \qquad (11)$$

where *ELT* is the number of events for the equipment's economic lifetime; *AOH* is the number of outcomes for annual operation hours; $P_{\{X=m\}}^{ELT}$ and $P_{\{X=n\}}^{AOH}$ are the probability of each event; and $C_i^h(m,n)$ is the hourly ownership cost for "*m*" hours of DiGs' useful lifetime and "*n*" hours of annual operation for the $i^{th}$ approach. From the expected value, it is possible to calculate the risk (standard deviation) for the four approaches discussed as follows:

$$\sigma_i = \sqrt{\left\{EV_i\left[X^2\right] - \left[EV_i\left[X\right]\right]^2\right\}} \qquad (12)$$

Using (11), equation (12) can be rewritten as:

$$\sigma_i = \sqrt{\left\{\sum_{m=1}^{ELT}\sum_{n=1}^{AOH}\left\langle P_{\{X=m\}}^{ELT} \cdot P_{\{X=n\}}^{AOH} \cdot \left\{C_i^h(m,n)\right\}^2\right\rangle - \left[EV\left[C_i^h\right]\right]^2\right\}} \qquad (13)$$

$\sigma_i$ shows the risk associated with the $i^{\text{th}}$ approach considering uncertainties. The best approach is the one with lower risk as long as its total payment throughout the project lifetime is not significantly different (less than 1% different) from the base-case (i.e., approach I).

## IV. SIMULATION STUDIES AND RESULTS

In this section, comprehensive simulation results are presented for a specific DiG to assess the effectiveness of the proposed framework in its ownership cost calculation, and to find the best approach in comparison with the base-case approach. DiG is used as an example in this study since it is the most common device in MGs. Simulation results are divided into two sub-sections. In the first sub-section, the accuracy of Approaches II, III.A, and III.B is verified in comparison with the base-case through a set of simulation studies. In the second sub-section, sensitivity, uncertainty and risk analyses are given for the approaches that pass the verification test. As it was explained in Section III.D, the appropriate DPDFs will also be defined for each uncertain parameter prior to risk analysis.

### A. Verification Study:

Approach I is known as the standard machinery cost calculation method [24]-[26]. Therefore, other approaches should yield similar (or close) accumulative payment for the same outcomes. If accumulative payment throughout the project lifetime obtained from an approach is not close enough (within 1%) to the one obtained from the base-case, it will be rejected. For each approach, a simulation study is carried out where a DiG unit is considered to have 20000 hours of economic lifetime and 5000 annual operation hours for different project lifetime (4, 8, 12, 16, 20 years). The DiG selected for this study is a 10 kW diesel Genset [27]. Other required data for the simulation study is reported in Table I.

TABLE I
OWNERSHIP COST CALCULATION DATA.

| DiG capital cost ($C_{pur,0}$), $ | DiG replacement cost ($C_{rep}$), $ | Inflation rate ($i_{\text{inf}}$), % | Interest rate ($i_{\text{int}}$), % |
|---|---|---|---|
| 6750 | 4725 | 1.5 | 3.5 |

Simulation results for the different approaches are shown in Fig. 2, where their corresponding error is compared to the base-case approach. Total accumulative payment for each approach for different length of project lifetime is depicted with bars. It is clear from Fig. 2 that the accumulative payment in Approach II is not close to the base-case. Similar observation can be made by the error curves, where Approach II gives the highest error (about 4.8%). The error in Approach III.A is always less than 1%, while Approach III.B is more accurate for smaller length of project lifetime (with an error of near zero) and a maximum error of around 1.2% for 20 years of lifetime. Therefore, Approach II, which gave an error of near 5%, failed the verification test and should not be used for such studies because of large error. The error in Approaches III.A and III.B are within an acceptable range which means that they can be used in place of Approach I without jeopardizing the accuracy of the ownership cost calculations.

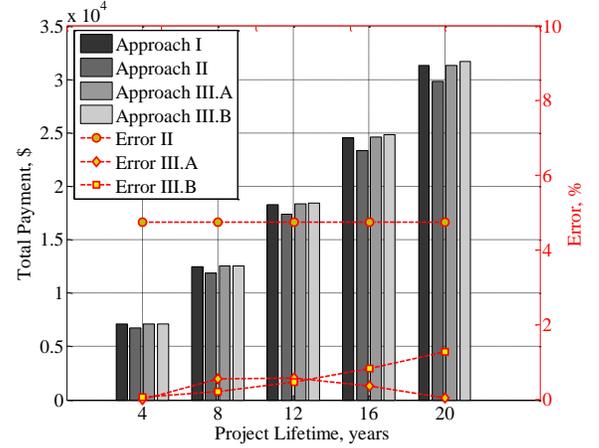

Fig. 2. Accumulative payment of the DiG ownership cost for different approaches and their corresponding error in comparison with Approach I.

Although the above simulation results give a perspective of the proposed Approaches, they do not represent a basis for comparison between different Approaches regarding uncertainty sources. Therefore, uncertainty and risk analyses are required for further evaluation of the proposed Approaches, given below.

### B. Uncertainty and Risk Analyses:

Prior to sensitivity and risk analyses, appropriate DPDFs are necessary to be assigned to each uncertainty source studied. Also, as mentioned earlier, two to five outcomes with discrete probabilities are required to be defined based on expert best judgment. These probabilities are obtained from the DPDFs.

*Appropriate DPDF for DiG useful lifetime*: DiGs usually work 15000-25000 hours depending on their rated power, quality and interval of maintenance, quality of diesel fuel, and loading condition throughout their lifetime [28]. In general with regular overhaul (i.e., complete maintenance), the typical lifetime of a DiG could be up to 13 years. It is also well-known that the economic lifetime is usually less for smaller DiGs. The 10-kW DiG under study is more likely to have an economic lifetime in the 17000-hour range. The Hypergeometric distribution is used to generate the probability values for the DiG economic lifetime. Hypergeometric distributions are originally used to describe samples where the selections from a binary set of items are not replaced [29]. In this study, the shape of the distribution is found to be appropriate where the probability of the lower values (with the highest probability at 17000 hours) is higher than the larger values, as shown in Fig. 3. The Hypergeometric PDF is given as follows [29]:

$$P^{ULT}_{\{X=k\}} = \frac{\binom{K}{k}\binom{N-K}{n-k}}{\binom{K}{n}} \quad (14)$$

where $N$ is the total number of items (70 in this study); $K$ is the total number of items with desired trait (14 in this study);

$n$ is the number of items in the sample (10 in this study); and $k$ is the number of items with desired trait in the sample (15000 to 20000 hours). The generated probabilities are shown in Fig. 3.

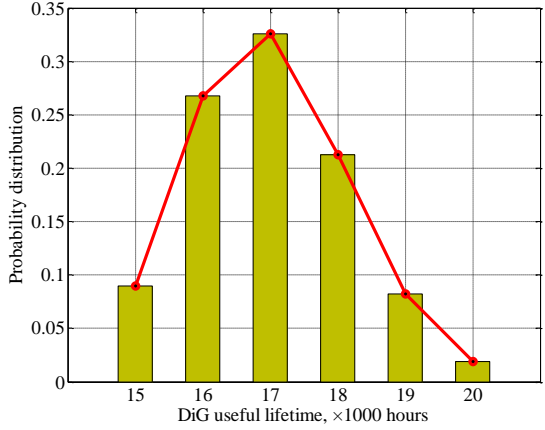

Fig. 3. PDF for DiG useful lifetime created by Hypergeometric DPDF.

*Appropriate DPDF for DiG annual operation hours*: In this case, another DPDF is required since the annual operation hours is a different source of uncertainty with a different behavior. In order to have an estimate of the DiG annual operation hours, HOMER® software [23] has been used to design an islanded MG for a remote area with three residential units each of which has a power consumption profile of a typical U.S. resident [30]. The component sizes and annual operation obtained for the MG (including PV solar panels, a DiG, and a battery bank), for a typical residential load, are given in Table II.

TABLE II
MG DESIGN SPECIFICATIONS OBTAINED FROM HOMER®.

|  | DiG | PV | Battery |
| --- | --- | --- | --- |
| Rated capacity | 10 kW | 17 kW | 24 kWh |
| Annual operation | 7676 hour/year | 3803 hour/year | 801 kWh/year |

Based on the MG specification given in Table II, it is estimated that the DiG will work around 7700 hours (out of the maximum 8760 hours) a year. However, it is possible that the DiG would operate more or less than 7700 hours per year, with a higher probability for larger annual operation hours. The reason for this can be two-fold: 1) HOMER software does not include the stochastic nature of renewable generation and load demand in the design procedure, and 2) the only renewable generation source in the designed MG is solar PV whose output power is unpredictable, which increases the required reserve that must be provided by the DiG. For this reason, the extreme value discrete distribution is selected as an appropriate DPDF for annual operation hours [31]:

$$P^{AOH}_{\{X|\mu,\sigma\}} = \sigma^{-1} \exp\left(\frac{X-\mu}{\sigma}\right) \cdot \exp\left(-\exp\left(\frac{X-\mu}{\sigma}\right)\right) \quad (15)$$

where it returns the PDF of type 1 extreme value distribution with location parameter $\mu$ (3 in this study) and scale parameter $\sigma$ (1.5 in this study), evaluated for the values in $X$. Extreme value distributions are frequently used to model the smallest or largest value among a large set of independent random values representing measurements or observations [31]. This DPDF is chosen since it produces the expected probability distributions following the hypothesis derived above for the annual operation hours. The probability distribution of the annual operation hours for the DiG is shown in Fig. 4.

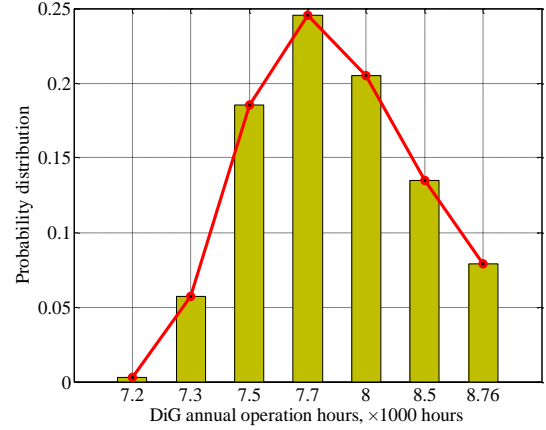

Fig. 4. PDF for DiG annual operation hours created by Extreme Value DPDF.

Although the DPDFs of the uncertainty sources can be different for different studies, similar analyses can be performed. Therefore, the proposed framework can be employed for different components of MGs. In this study, similar DPDFs will be used for different Approaches for fair comparison.

With different outcomes considered for the DiG annual operation hours and useful lifetime, it is possible to perform a sensitivity analysis by calculating hourly cost for each approach. The results from Approaches I, III.A and III.B are shown in Fig. 5(a), (b), and (c), respectively. The project lifetime is assumed to be 20 years in these simulations. The other parameters used in the simulation are given in Table I.

It can be seen from Fig. 5(b) that Approach III.A is more sensitive to the uncertain parameters (compared to the base-case in Fig. 5(a)); however its variations are smaller. Furthermore, Fig. 5(c) reveals that the variations in the annual operation hours have negligible impact on the ownership cost (less than 1%) in Approach III.B. This is a significant improvement in the uncertainty analysis.

The expected ownership cost and risk value (i.e., standard deviation) for Approaches I, III.A, and III.B are calculated for the DiG using (11)-(13). The results are given in Table III. Although the expected ownership cost is not meaningful in terms of uncertainty and risk analyses, it provides the hourly ownership cost which has to be included in the operational cost of the DiG. Alternatively, looking at the risk value associated with each Approach shows that Approach III.B has the lowest risk in the calculation in the presence of uncertainty. In other words, it basically shows that Approach III.B is significantly less risky (about 22% compared to Approach I), considering possible variations in the uncertain parameters, when calculating ownership cost. Therefore, it can be concluded that Approach III.B is the best method (among the four approaches) to calculate the ownership cost with high accuracy and lower risk associated with the two uncertainty sources.

however sometimes slight modifications may be required to make the framework fit a specific DER. Specifically, one special modification is required when the useful lifetime of the device is expressed in something other than hours. For instance, battery useful lifetime is expressed as battery nominal charge life (A.h.). Therefore, it is essential to modify the equations to adapt them to batteries. This section covers such modifications needed to calculate the ownership cost of batteries.

As it was discussed in Section III, the useful lifetime of battery and its salvage value are required for depreciation cost calculation. The economic life of battery banks is the number of years (in terms of battery A.h.) for which costs are to be estimated. This is important, particularly for Approaches III.A and III.B, where the economic life of the battery bank with the required number of replacement(s) throughout the project lifetime should be considered. A battery has a finite lifespan which is quantified by summing together the entire A.h. discharged during each of its cycles to determine a cumulative A.h. capacity for the battery's life. The total charge life of a battery is calculated from manufacturer's specifications as follows [14], [18]:

$$\Gamma_r = L_r \cdot D_r \cdot C_r \qquad (16)$$

where $\Gamma_r$ is the charge life of a battery (A.h.); $L_r$ is its cycle lifetime at rated depth of discharge (DoD) and discharge current; $D_r$ is the rated DoD; and $C_r$ is the A.h. capacity at rated discharge current. In reality, a battery is not always operated at its rated values, and it is well-known that the total A.h. through a battery during its effective lifetime is directly dependent upon the DoD during each cycle [14], [18]. Therefore, a relationship can be established between the cycle-to-cycle DoD and battery effective lifetime. The unified effects of DoD and discharge rate on battery lifespan can be expressed as follows [18]:

$$d_{eff}(i) = d_{act}(i) \cdot \left(\frac{C_r}{C_{act}(i)}\right) \cdot \left(\frac{DoD_{act}(i)}{DoD_r}\right)^{u_0} \cdot e^{u_1\left(\frac{DoD_{act}(i)}{DoD_r}-1\right)} \qquad (17)$$

where $d_{eff}(i)$ is the effective A.h. discharge of the battery at discharge event $i$; $d_{act}(i)$ is the actual discharged energy of the battery (A.h.); $C_r$ is the battery A.h. capacity at rated discharge current; $C_{act}$ is the actual battery capacity at each time interval; $DoD_{act}(i)$ is the actual DoD at discharge event $i$ (%); $DoD_r$ is the rated DoD (%) in event $i$; and $u_0$ and $u_1$ are curve-fitting coefficients. All the equations for calculating the depreciation cost in each Approach are valid here as well. For example, to obtain the battery ownership cost using Approach III.B, (10) is modified to calculate the ownership cost per effective A.h. by estimating the annual accumulative discharged A.h. as follows:

$$C_{III.B}^{A.h} = \frac{C_{dep}}{\sum_{each\ year} d_{eff}(i) \times (NO_{rep}+1)} \times (1+i)^j \qquad (18)$$

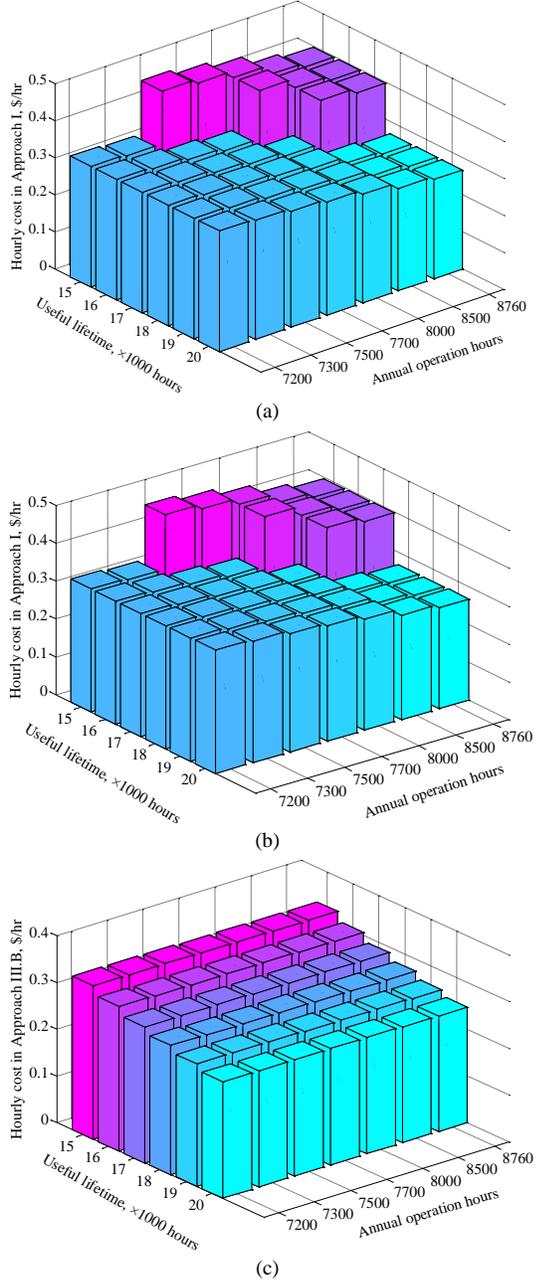

Fig. 5. Hourly cost for the DiG with different outcomes generated for the two uncertain parameters for (a) Approach I (base-case), (b) Approach III.A, and (c) Approach III.B.

TABLE III
OWNERSHIP COST CALCULATION DATA FOR THE DiG.

| Terms of analysis | Approach I | Approach III.A | Approach III.B |
|---|---|---|---|
| Expected Cost ($/hour) | 0.3042 | 0.3377 | 0.2644 |
| Risk Value ($/hour) | 0.1113 | 0.1111 | 0.0869 |

In the next section, the formulation of ownership cost for storage battery as another DER in a MG setting is presented.

## V. OWNERSHIP COST FORMULATION FOR BATTERIES

Generally speaking, the proposed framework for ownership cost calculation can be applicable to other MG components,

The uncertainty and risk analyses for this situation can be achieved by following the steps given in Section III. Note that appropriate DPDFs for a battery might be different from those given for DiG and are based on the annual usage and useful lifetime of the battery.

Simulation studies have been carried out for battery and yield similar results as for DiG, where Approach III.B was found to be the best model.

## VI. Conclusion

Cost-based operation of DERs and MGs, including their ownership costs, is inevitable in the smart grid era. This paper proposes a general framework for ownership cost calculation considering uncertainty and risk associated with the calculations. Two new ownership cost calculation approaches are proposed and compared with two approaches from the literature. It is shown through a series of simulation studies that the new Approaches III.A and III.B yield very accurate results in comparison with the base-case (Approach I), which is commonly used in the literature. In addition, a risk analysis of the proposed approaches shows that the proposed Approach III.B is 22% less risky compared to Approach I. It is also shown through sensitivity analysis that Approach III.B is more robust against the annual operation hour variations, unlike the other Approaches discussed.

In general, the proposed ownership cost calculation framework can be utilized (with minor modifications) for different components of MGs. As an example, in addition to DiG, the proposed framework is modified and given for batteries. Simulation results for battery (not given in this paper) also reached the same conclusion that Approach III.B is the most robust one.